\documentclass[twocolumn,superscriptaddress,showpacs,aps,floatfix]{revtex4-1}
\usepackage{amsmath,amsfonts,amssymb,epsfig,dcolumn,bm,dsfont,graphics,latexsym,color,graphicx}

\let\oldAA\AA

\renewcommand{\AA}{\text{\normalfont\oldAA}}
\newcommand{\ket}[1]{| {#1} \rangle} 
\newcommand{\bra}[1]{\langle {#1} |} 

\begin{document}
\preprint{AIP/123-QED}
\title{Control of a spin qubit in a lateral GaAs quantum dot based on symmetry of gating potential}
\author{Pavle Stipsi\'c}
\affiliation {Faculty of Physics, University of Belgrade, Studentski trg 12, 11001 Belgrade, Serbia}
\author{Marko Milivojevi\'c}
\affiliation {NanoLab, QTP Center, Faculty of Physics, University of Belgrade, Studentski trg 12, 11001 Belgrade, Serbia}
\affiliation{Department of Theoretical Physics and Astrophysics, Faculty of Science,
P. J. \v{S}af\'{a}rik University, Park Angelinum 9, 040 01
Ko\v{s}ice, Slovak Republic}
\begin{abstract}
We study the influence of quantum dot symmetry on the Rabi frequency and phonon-induced spin relaxation rate in a single-electron GaAs spin qubit.
We find that anisotropic dependence on the magnetic field direction is independent of the choice of the gating potential.
Also, we discover that relative orientation of the quantum dot, with respect to the crystallographic frame,
is relevant in systems with ${\bf C}_{1{\rm v}}$, ${\bf C}_{2{\rm v}}$, or ${\bf C}_{n}$ ($n\neq4r$) symmetry.
To demonstrate the important impact of the gating potential shape on the spin qubit lifetime,
we compare the effects of an infinite-wall equilateral triangle,
square, and rectangular confinement with the known results for the harmonic potential.
In the studied cases, enhanced spin qubit lifetime is revealed, reaching almost six
orders of magnitude increase for the equilateral triangle gating.
\newline\newline
PACS numbers:
81.07.Ta, 
71.70.Ej, 
72.10.Di, 
76.30.−v, 
31.15.Hz 
\end{abstract}
\maketitle
\section{Introduction}
Every quantum two-level system can act as the
quantum bit, a basic unit of quantum information processing~\cite{NC10,BdV}.
Among different solid-state implementations of the qubit system~\cite{LdV98,OTW+99,NPT99,YTB+12},
single-electron spin in a semiconductor quantum dot (QD)
can be used to achieve the task. In order to manipulate
spins of charge carriers embedded inside a semiconductor
material electrically, through electric dipole spin resonance
(EDSR)~\cite{R60}, the presence of spin-orbit interaction (SOI) is obligatory.

Besides its positive effect in EDSR based schemes~\cite{GBL06,BL07,NKN+07,R08,BSO+11,KSW+14,TYO+18,KLS19,SKT+19},
SOI enables the electron-phonon coupling
mediated transitions between the qubit states~\cite{KN00,PF05,S17,LLH+18},
affecting the spin qubit lifetime.
To suppress the coupling to phonons, different approaches
like the optimization of the QD design~\cite{CBG+07,MSL16} or
control of the system size~\cite{CM07} were suggested.
The observed anisotropy of the spin relaxation rate
on the in-plane magnetic field orientation~\cite{FAT05}
offered another playground for fine-tuning
of the spin qubit's desired properties.
In circular QDs, this is the only degree of freedom
accessible in the optimization of the spin qubit,
while for the elliptical confining potential~\cite{SKS+14,MSL16,OS07,AMR+08}
orientation of the QD potential with
respect to the crystallographic frame can be used as the tuning parameter.

Evidently, different symmetry of the
gating potential~\cite{PRC15} is the main reason
for the observed behavior. But to what extent
can the potential symmetry alter the basic
properties of the electrically controlled spin qubit?
To address this question, we have performed
a general analysis valid for the lateral GaAs QD system
with ${\bf C}_{n{\rm v}}$ or ${\bf C}_{n}$ symmetry of the gating potential.
Besides the expected anisotropy on the
magnetic field orientation, we were able to
find potential symmetries for which the QD
orientation with respect to the crystallographic
frame can act as another control parameter
of the spin qubit characteristics.
With our theory, we offer a simple and
efficient way to determine the impact
of the gating potential on the Rabi
frequency and spin relaxation rate. This is
shown in the example of anisotropic and
isotropic harmonic potential, as well as
for the infinite-wall equilateral triangle, square, and
rectangular potential.

This paper is organized as follows.
In Section~\ref{Section2} we
define a single-electron GaAs spin qubit model.
In Section~\ref{Section3}
we define the dipole moment
of the electrically controlled spin qubit
that describes both the Rabi frequency
and SOI-induced spin relaxation rate mediated by acoustic phonons.
In Section~\ref{Dipole} we present the main results of the paper:
analytical expressions for the dipole moment
in the case of the gating potential with ${\bf C}_{n{\rm v}}$ or
${\bf C}_{n}$ symmetry.
In Section~\ref{examples}, to illustrate
the impact of the gating potential
on the spin qubit lifetime
we use the obtained expressions
to compare the influence of the
harmonic confinement
with an infinite-wall equilateral triangle,
square, and rectangular potential.
In Section~\ref{Section4} we give our conclusions.

\section{Dynamics of the lateral QD}\label{Section2}
\begin{figure}[t]
\includegraphics[width=8.5cm]{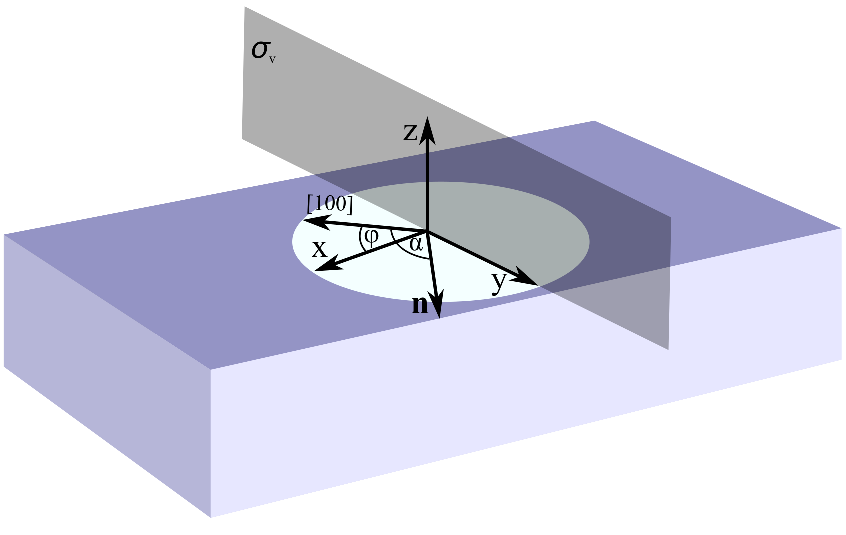}
\caption{\label{setup} A schematic view of the GaAs lateral QD.
The $y$ axis of the QD reference frame coincides with the vertical mirror plane symmetry $\sigma_{\rm v}$.
We define the angle between the chosen $x$ axis and the crystallographic $[100]$ axis as $\varphi$.
The magnetic field is aligned along the ${\bf n}$ direction,
forming an angle $\alpha$ with the $[100]$ direction.}
\end{figure}

We start with the Hamiltonian describing the lateral dynamics of a single-electron in the GaAs material
\begin{equation}\label{Ham}
  H=H_0+H_{\rm z}+H_{\rm so}=\frac{p_x^2+p_y^2}{2m^*}+V(x,y)+H_{\rm z}+H_{\rm so},
\end{equation}
where $p_x$ and $p_y$ are the momentum operators,
$m^*$ is the effective mass ($m^{*}=0.067 m_e$ for GaAs, $m_e$ is the electron mass), while
$V(x,y)$ is the gating potential used to localize the electron in a QD.
In the lateral system, symmetries that can be present are
the $n$-fold rotational symmetry and the vertical mirror plane symmetry $\sigma_{\rm v}$.
For simplicity, we assume that $\sigma_{\rm v}$ coincides with the $yz$ plane
of the QD coordinate frame (see FIG.~\ref{setup}).
Thus, we assume a general form of the orbital Hamiltonian
$H_0$ that has a ${\bf C}_{n{\rm v}}$ or ${\bf C}_{n}$ ($n=\infty$ also) symmetry.
Due to the symmetry, eigenenergies and eigenvectors of $H_0$ can be classified according to
the irreducible representations (IRs) of a given point-group symmetry.

Besides $H_0$, in Eq.~\eqref{Ham} the Zeeman term $H_{\rm z}$ appears, describing the coupling of spin and
magnetic field:
\begin{equation}\label{zeman}
  H_{\rm z}=g\mu_B {\bm B}\cdot{\bm s},
\end{equation}
where $g$ is the effective Land\'{e} factor ($g\approx-0.44$ for GaAs),
$\mu_B$ is the Bohr magneton, ${\bm s}=1/2 {\bm \sigma}$ is the electron's spin,
and ${\bm B}=B{\bm n}$ is the in-plane magnetic field
forming an angle $\alpha$ with the crystallographic [100] axis.
In Eq.~\eqref{Ham} we have neglected the orbital effects of
the in-plane magnetic field. This is a reasonable assumption
for the magnetic field strength weaker than a few ${\rm T}$~\cite{RSF11}.
In the case of the magnetic field applied in the $z$ direction, orbital effects would be much more pronounced~\cite{RSF11}.

Eigenstates of $H_0+H_{\rm z}$ can be written in a direct product
form $\ket{\Psi_i\pm}=\ket{\Psi_i}\otimes{\ket{\pm}}$, where $\ket{\Psi_i}$ corresponds to the
eigenvectors of the Hamiltonian $H_0$ with an energy $\epsilon_i$, while $\ket{\pm}$
represents eigenvectors of $H_{\rm z}$ with spin projection parallel or antiparallel
to the magnetic field direction and an eigenenergy $\pm g\mu_BB/2$, respectively.
The effect of $H_{\rm z}$ on the eigenspectra of $H_0$ can be seen as the splitting of $H_0$ eigenenergies
into two branches with an energy difference $|g|\mu_B B$. In this work, we assume that $|g|\mu_B B$ is much weaker than
the energy difference between the ground and the first excited state of the orbital Hamiltonian $H_0$.

Besides $H_0$ ($H_{\rm z}$) that acts trivially in the spin (orbital) space,
the SOI Hamiltonian does not commute with $H_0+H_{\rm z}$.
It consists of two terms, Dresselhaus~\cite{DR} and Rashba~\cite{RA}:
the Dresselhaus term exists due to the bulk inversion asymmetry of the structure,
while the Rashba term is present when an electric field
perpendicular to the growth direction is applied.
The form of spin-orbit coupling is dependent on the structure's symmetry.
For GaAs, having the zincblende structure,
the SOI Hamiltonian is equal to
\begin{eqnarray}
  H_{\rm so}= 2\alpha_{\rm d}(p_{y}^{\rm c}s_y-p_{x}^{\rm c}s_x)+2\alpha_{\rm r}(p_{x}^{\rm c}s_y-p_{y}^{\rm c}s_x),\;\;\label{prva}
\end{eqnarray}
where $\alpha_{\rm r}$ and $\alpha_{\rm d}$ are Rashba and Dresselhaus coupling constants,
while $p_{x}^{\rm c}$ and $p_{y}^{\rm c}$ are momentum operators in the [100] and [010] crystallographic directions, respectively.
The electron spin is locked to the crystal momentum, since the potential trap confines electron of the crystal.
Thus, an electron in a QD inherits the features of the crystal for which the crystal momentum is only appropriately defined.
However, we have the choice to define the $x$ axis of our coordinate frame independently on the crystallographic [100] direction.
Assuming that the angle between them is $\varphi$, $p_{x}^{\rm c}$ and $p_{y}^{\rm c}$ should be written in terms of momentum operators in the
chosen frame: $p_{x}^{\rm c}=p_x \cos{\varphi}-p_y\sin{\varphi}$, $p_{y}^{\rm c}=p_x\sin{\varphi}+p_y\cos{\varphi}$.

The spin-orbit Hamiltonian can be written in a different form using the  Rashba $l_{\rm r}=\hbar^2/2m\alpha_r$ and Dresselhaus $l_{\rm d}=\hbar^2/2m\alpha_d$  precession lengths,
\begin{eqnarray}
  H_{\rm so}= \hbar\Big(\frac{p_{y}^{\rm c}s_y-p_{x}^{\rm c}s_x}{m^* l_{\rm d}}+\frac{p_{x}^{\rm c}s_y-p_{y}^{\rm c}s_x}{m^* l_{\rm r}}\Big).\label{druga}
\end{eqnarray}
To compare the ratio of the spin-orbit precession length and the orbital confinement length
$l$, we redefine $l_{\rm r}$ and $l_{\rm d}$ in terms of the overall
spin-orbit length $l_{\rm so}$ and the spin-orbit angle $\nu$:
\begin{equation}\label{zz}
  l_{\rm d}^{-1}=l_{\rm so}^{-1}\sin{\nu},\;l_{\rm r}^{-1}=l_{\rm so}^{-1}\cos{\nu}.
\end{equation}
Since we assume no doping of the GaAs material~\cite{SL05}, $l_{\rm so}$ can be considered constant.
Moreover, the relation $l_{\rm so}\gg l$~\cite{RSF11,RPS+14} is satisfied in GaAs QDs,
meaning that SOI can be treated as a perturbation.

Without SOI, qubit states can be defined as $\ket{\Psi_0\pm}=\ket{\Psi_0}\otimes\ket{\pm}$,
where $\ket{\Psi_0}$ corresponds to the ground state of the spin-independent Hamiltonian $H_0$.
Because SOI can be treated on the level of a perturbation,
we calculate first-order corrections of the qubit states due to
spin-orbit coupling.
Since it is known that the standard perturbation technique badly incorporates the spin-orbit-induced corrections~\cite{CWL04,BSF10},
we follow the procedure explained in Ref.~\cite{MSL16}: the Hamiltonian $H$ is transformed
using the unitary operator $U={\rm exp}({\rm i}{\bf n}_{\rm so}\cdot{\bm s})$, defined with the help of the position-dependent spin-orbit vector
${\bf n}_{\rm so}=l_{\rm so}^{-1}(r_{1}\sin{\nu}+r_{2}\cos{\nu},-r_{1}\cos{\nu}-r_{2}\sin{\nu},0)$:
\begin{equation}\label{Sch-Wolf}
  UHU^{\dag}=H_{0}+H_{\rm z}+H_{\rm so}^{\rm eff}.
\end{equation}
The unitary operator $U$ does not change the orbital and Zeeman Hamiltonian.
On the other hand, the SOI Hamiltonian $H_{\rm so}$ is transformed into
\begin{equation}\label{SOeff}
  H_{\rm so}^{\rm eff}=g\mu_B({\bf n}_{\rm so}\times {\bf B})\cdot{\bm s}-\frac{\hbar^2}{4m^*l_{\rm so}^2}
  \Big(1+2l_z s_z\cos{2\nu}\Big),
\end{equation}
where $l_z=-{\rm i}(r_{1}\partial_{r_{2}}-r_{2}\partial_{r_{1}})$ is the orbital angular momentum.
Using $H_{\rm so}^{\rm eff}$, the first-order correction of the qubit states can be written as
\begin{equation}
  \delta\ket{\Psi_{0}\sigma'} = U\sum_{i\neq0,\sigma^{\prime\prime}}\frac{\bra{\Psi_{i}\sigma^{\prime\prime}}H_{\rm so}^{\rm eff}\ket{\Psi_{0}\sigma^{\prime}}}{\epsilon_{0}-\epsilon_{i}+\frac{\sigma^{\prime}-\sigma^{\prime\prime}}2 g\mu_B B}\ket{\Psi_{i}\sigma^{\prime\prime}},
\end{equation}
where the sum over $i\neq0$ corresponds to all orbital eigenvectors $\ket{\Psi_{i}}$ different from the ground state $\ket{\Psi_{0}}$,
while $\sigma^{\prime\prime}=\pm$.

The lateral QD model is valid if the electron dynamics in the $z$ direction is suppressed; i.e., an electron is always in the ground state.
Thus, we assume that confinement length in the $z$ direction is much stronger than in the $xy$ plane.
The Hamiltonian describing the quantum confinement in the $z$ direction is equal to
$H(z)=p_z^2/2m^*+V(z)$, where $V(z)=eE_0z$ for $z\ge0$ and $V(z)=\infty$ for $z<0$.
To this Hamiltonian corresponds the following ground state (for $z>0$)~\cite{SS03}
\begin{equation}\label{zz}
  \Psi_0(z)=1.4261 \sqrt{\chi}{\rm Ai}(\chi z-2.3381),
\end{equation}
where ${\rm Ai}$ is the Airy function,
while $\chi=(2m^*eE_0/\hbar^2)^{1/3}$ is the inverse of
the characteristic length $z_0=1.5587/\chi$ in the $z$ direction.

In order to simplify the notation, in the rest of the paper
we assume that $\ket{\uparrow}$ and $\ket{\downarrow}$
represent SOI corrected qubit states in the $xy$-plane,
while $\ket{\Psi_{\uparrow}}=\ket{\uparrow}\Psi_0(z)$ and
$\ket{\Psi_{\downarrow}}=\ket{\downarrow}\Psi_0(z)$
correspond to wavefunctions of the qubit states in three
dimensions.

\section{Rabi frequency and phonon induced spin relaxation rate}\label{Section3}
Electrical control of the spin qubit is possible
by applying the in-plane oscillating electric field ${\bf E}\cos{\omega t}$,
resulting in the Rabi Hamiltonian $H_{\rm R}=e {\bf E}\cdot{\bf r}\cos(\omega t)$.
The Rabi frequency, measuring the speed of the single-qubit rotations, is equal to
$\Omega=e/\hbar|{\bf E}\cdot\bra{\uparrow}{\bf r}\ket{\downarrow}|$,
where
\begin{equation}\label{Rabi-freq}
{\bf d}_{\uparrow\downarrow}=\bra{\uparrow}{\bf r}\ket{\downarrow}
\end{equation}
is the dipole moment (in $e$ units), present due to the SOI induced spin mixing mechanism.
Misalignment of the applied field direction and the dipole moment leads to
a trivial suppression of the Rabi frequency.
Since it is beneficial to increase the Rabi frequency as much as possible,
the electric field should be applied in the direction of the dipole moment.
Thus, for fixed $|{\bf E}|$, the maximal value $\max(\Omega)=\Omega_{\uparrow\downarrow}$ of the Rabi frequency
\begin{equation}\label{Rabi-freq}
  \Omega_{\uparrow\downarrow}=\frac{e}{\hbar}|{\bf E}||{\bf d}_{\uparrow\downarrow}|
\end{equation}
is completely dependent on the strength of the dipole moment.

Since spin-phonon interaction in semiconductor QDs is irrelevant~\cite{KN00}, unlike donor-bound electrons in direct band-gap semiconductors~\cite{LKD+16},
only electron-phonon-induced transition between the qubit states should be considered in the study of spin relaxation.
Electron-phonon coupling is triggered by the SOI-induced admixture mechanism, being highly dependent on the
symmetry of the gating potential~\cite{LKD+16}.
We determine the rate of spin relaxation at $T=0$ from the Fermi golden rule,
\begin{equation}\label{FGR}
  \Gamma_{\uparrow\downarrow}=\frac{2\pi}{\hbar}\sum_{\nu{\bf q}}|M_{\nu}({\bf q})|^2
  |\bra{\Psi_{\uparrow}}{\rm e}^{{\rm i}{\bf q}\cdot{\bf r}_{\rm c}}\ket{\Psi_{\downarrow}}|^2\delta(\epsilon_{\uparrow\downarrow}-\hbar \omega_{\nu{\bf q}}),
\end{equation}
assuming the dominant contribution of acoustic phonons, having an energy $\hbar \omega_{\nu{\bf q}}$, equal to the level separation between the qubit states, $\epsilon_{\uparrow\downarrow}=|g|\mu_B B$.
For magnetic field strengths up to a few T, relevant for this work, the
linear dependence of phonon frequencies on the crystal wave vector length can be used, $\omega_{\nu{\bf q}}=c_{\nu}|{\bf q}|$,
giving us $|{\bf q}|=|g|\mu_B B/\hbar c_{\nu}$~\cite{coorframe}.

The geometric factor $|M_{\nu}({\bf q})|^2$ is dependent on the phonon mode,
longitudinal (LA) or transverse (TA).
The longitudinal geometric factor~\cite{CBG+06}
\begin{equation}\label{la}
|M_{\rm LA}({\bf q})|^2=\frac{\hbar D^2}{2\rho c_{{\rm LA}}V}|{\bf q}|+
\frac{32\pi^2\hbar (eh_{14})^2}{\epsilon^2\rho c_{{\rm LA}}V}\frac{(3q_xq_yq_z)^2}{|{\bf q}|^7}
\end{equation}
depends on both $D$ and $h_{14}$, representing the deformation and piezoelectric constant, respectively.
On the other hand, the transverse geometric factor~\cite{CBG+06}
\begin{eqnarray}\label{ta}
|M_{\rm TA}({\bf q})|^2&=&2\frac{32\pi^2\hbar (eh_{14})^2}{\epsilon^2\rho c_{{\rm TA}}V}\\
&&\times\left|\frac{q_x^2q_y^2+q_x^2q_z^2+q_y^2q_z^2}{|{\bf q}|^5}-\frac{(3q_xq_yq_z)^2}{|{\bf q}|^7}\right|\nonumber
\end{eqnarray}
is dependent on the piezoelectric constant solely.
Other parameters for the GaAs material are~\cite{CWL04,MSL16} $c_{\rm LA}=5290 {\rm m}/{\rm s}$, $c_{\rm TA}=2480 {\rm m}/{\rm s}$,
$\rho=5300{\rm kg}/{\rm m^3}$, $D=7{\rm eV}$, $eh_{14}=1.4\times10^9 {\rm eV}/{\rm m}$, and $\epsilon=12.9$.

Finally, in Eq.~\eqref{FGR} both the lateral and the $z$-direction confinement enter the relaxation rate
through the scattering matrix element $|\bra{\Psi_{\uparrow}}{\rm e}^{{\rm i}{\bf q}\cdot{\bf r}_{\rm c}}\ket{\Psi_{\downarrow}}|^2$.
We employ the dipole approximation ${\rm e}^{{\rm i}{\bf q}\cdot {\bf r}_c}\approx 1+{\rm i}{\bf q}\cdot {\bf r}_c$,
justified for magnetic field strengths below a few T.

To summarize, the phonon-induced relaxation rate can be divided into three separate channels:
the deformation phonons $\Gamma_{\uparrow\downarrow}^{\rm def}$,
the longitudinal piezoelectric phonons $\Gamma_{\uparrow\downarrow}^{\rm piez,LA}$,
and the transverse piezoelectric phonons $\Gamma_{\uparrow\downarrow}^{\rm piez,TA}$.
In GaAs QDs, $ \Gamma_{\uparrow\downarrow}^{\rm piez,TA}$ is the
dominant relaxation channel, being two orders of magnitude
stronger than  $\Gamma_{\uparrow\downarrow}^{\rm piez,LA}+\Gamma_{\uparrow\downarrow}^{\rm def}$
in the dipole approximation regime.
Thus, we can identify the total relaxation rate with $\Gamma_{\uparrow\downarrow}^{\rm piez,TA}$~\cite{CSZ+18}:
\begin{eqnarray}\label{v}
     \Gamma_{\uparrow\downarrow}&=&\frac{256\pi(eh_{14})^2(|g|\mu_BB)^3}{105c_{\rm TA}^5\rho\hbar^4\epsilon^2}(1+\frac{7}{33}K_{\rm TA}^2z_0^2)
    |{\bf d}_{\uparrow\downarrow}|^2\label{relaxFORMULA3},\;\;\;\;\;\;
  \end{eqnarray}
where $K_{\rm TA}=|g|\mu_BB/\hbar c_{\rm TA}$.
We assume a typical confinement length $l=10$nm~\cite{RSF11,RPS+14} of the GaAs QD in an experimental setup and magnetic field up to a few T (see Section~\ref{Section2}). Since confinement in the $z$ direction is much stronger than in the $xy$ plane, $z_0\ll l$, we conclude that
$7K_{\rm TA}^2z_0^2/33$ is much weaker than 1. In other words, the influence of the confinement in the $z$ direction can be neglected.

Note that $\Gamma_{\uparrow\downarrow}$ is squarely dependent on the
absolute value of the dipole moment, meaning that the
knowledge of the dipole moment is sufficient to fully
explain the behavior of both the Rabi frequency and the spin relaxation rate.
\begin{figure}[t]
\includegraphics[width=8.5cm]{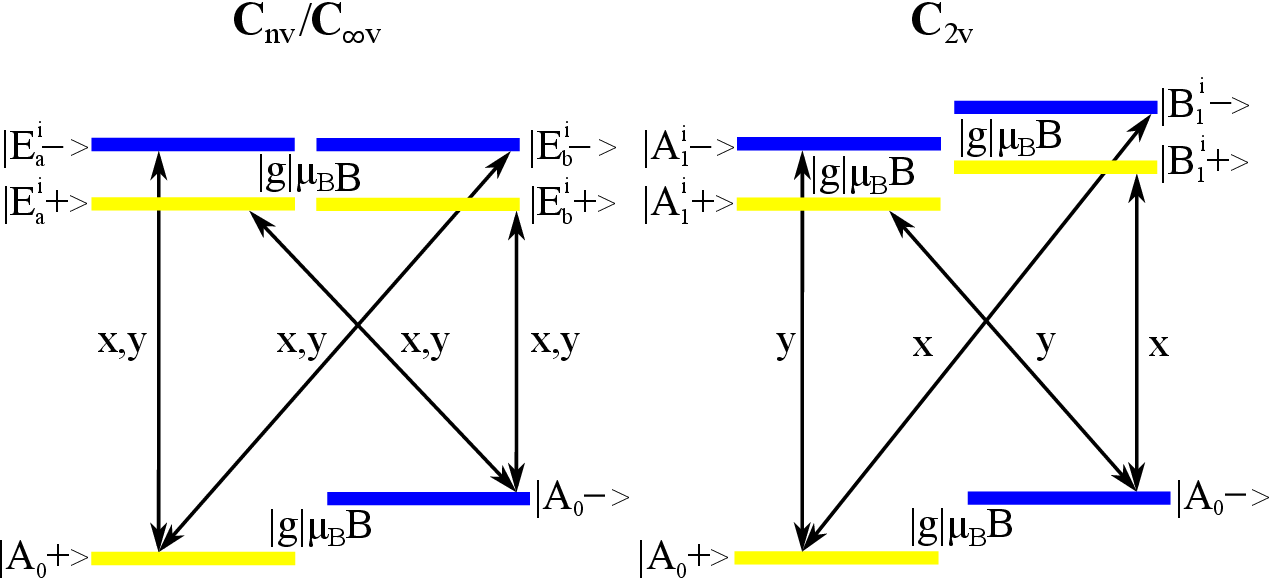}
\caption{\label{SCHEME} A schematic view of the first-order perturbation correction of the
qubit states $\ket{A_0\pm}$ in the case of ${\bf C}_{n{\rm v}}({\bf C}_{\infty{\rm v}})$ (left)
and ${\bf C}_{2{\rm v}}$ (right) symmetry.
In the first case, states that correct the qubit states
have twofold orbital degeneracy and transform according
to the IR $E_1$. These states are split by the Zeeman energy $|g|\mu_B B$.
The transition between the SOI uncorrected qubit states and the $\ket{E_{a,b}^i}$ states is enabled by the $x$ and $y$ terms from $H_{\rm so}^{\rm eff}$.
In the second case, orbital states involved in the qubit states correction transform according
to IRs $A_1$ and $B_1$; the transition is triggered by
the terms $y$ and $x$ from $H_{\rm so}^{\rm eff}$, respectively.}
\end{figure}
\section{Analytical expression for the dipole moment}\label{Dipole}
Based on the previous conclusion, we come to the main objective: to derive symmetry
allowed expression for the dipole moment.
The results can be divided into three cases, according to the system's group symmetry:
(1) ${\bf C}_{n{\rm v}}$ ($n\geq3$) and ${\bf C}_{\infty {\rm v}}$,
(2) ${\bf C}_{2{\rm v}}$ and ${\bf C}_{1{\rm v}}$, and
(3) ${\bf C}_{n}$ and ${\bf C}_{\infty}$.

\subsection{Dipole moment for systems with ${\bf C}_{n{\rm v}}$ ($n\geq3$) or ${\bf C}_{\infty {\rm v}}$ symmetry}
To find the SOI-induced perturbative correction of the qubit states,
we first rewrite the unitarily transformed SOI Hamiltonian
in the coordinate frame of the potential,
\begin{eqnarray}\label{soieff}
  H_{\rm so}^{\rm eff}&=&g\mu_B B s_z\Big(x
    \big(\sin{(\nu+\varphi)}\sin{\alpha}+\cos{(\nu-\varphi)}\cos{\alpha}\big)\nonumber\\
&&+y\big(\cos{(\nu+\varphi)}\sin{\alpha}+\sin{(\nu-\varphi)}\cos{\alpha}\big)\Big),\label{hsoeff}
\end{eqnarray}
and neglect the second term in Eq.~\eqref{SOeff},
assuming magnetic field strengths $>\mu$T
needed to appropriately define the qubit states.
For simplicity, we define two factors,
\begin{eqnarray}
  v_x &=& \sin(\nu+\varphi)\sin{\alpha}+\cos(\nu-\varphi)\cos{\alpha}, \\
  v_y &=& \cos{(\nu+\varphi)}\sin{\alpha}+\sin{(\nu-\varphi)}\cos{\alpha},
\end{eqnarray}
with whose help $H_{\rm so}^{\rm eff}$ can be written in a more compact form.

The Hamiltonian $H_{\rm so}^{\rm eff}$ is in the orbital space dependent on the coordinates
$x$ and $y$ that transform according to the IR $E_1$.
Their symmetry behavior restricts the states
that can appear in the perturbative correction of the
qubit states. It is simple to check that only
states transforming according to the IR $E_1$ are allowed.
This is illustrated in the left-hand panel of FIG.~\ref{SCHEME}.

We label the ground state of the orbital Hamiltonian as $\ket{A_0}$,
since the ground state in quantum mechanical systems is of the
maximal possible symmetry~\cite{LCC08} and it should transform
according to the $A_0$ IR, representing the objects
invariant under all group symmetry operations (see Table~\ref{Cnv}).
We write two complex conjugate basis vectors of the two-dimensional IR $E_1$
as $\ket{E_a^i}$ and $\ket{E_b^i}$, where $i$ labels the energy level.
Also, we define the energy difference between the excited
level and the ground state as $\epsilon^i=\epsilon_{ex}^i-\epsilon_{gr}$.
\begin{table}[t]
\caption{\label{Cnv} For ${\bf C}_{n{\rm v}}$ and ${\bf C}_{\infty{\rm v}}$ symmetry groups, tables of matrices of the corresponding IRs are given~\cite{JB67},
tabulated on the generators $C_n$($R_{\beta}$) and $\sigma_{\rm v}$,
where $C_n$($R_{\beta}$) represents a rotation for the angle $2\pi/n$ ($\beta$)
around the $z$ axis.
In the ${\bf C}_{n{\rm v}}$ case, two-dimensional IRs exist if $n\geq3$.
In both cases, two-dimensional IRs are written in a complex conjugate basis.}
\begin{tabular}{l@{\hspace{4mm}}l@{\hspace{4mm}}l@{\hspace{4mm}}l@{\hspace{4mm}}l@{\hspace{1mm}}}\hline\noalign{\smallskip}
${\bf C}_{n{\rm v}}$&IR&$m$&${C}_n$&$\sigma_{\rm v}$\\ \noalign{\smallskip}\hline\noalign{\smallskip}
&$A_0/B_0$&0 &$1$&$\pm1$\\ \noalign{\smallskip}\hline\noalign{\smallskip}
&$E_m$& $(0,\frac{n}2)$&$\left(
                         \begin{array}{cc}
                           {\rm e}^{{\rm i}\frac{2\pi}n m} & 0 \\
                           0 & {\rm e}^{-{\rm i}\frac{2\pi}n m} \\
                         \end{array}
                       \right)$
&$\left(\begin{array}{cc}
                           0 & 1 \\
                           1 & 0 \\
                         \end{array}
                       \right)$\\ \noalign{\smallskip}\hline\noalign{\smallskip}
&$A_{\frac{n}2}/B_{\frac{n}2}$& $\frac{n}2$  &-1&$\pm1$\\ \noalign{\smallskip}\hline\hline\noalign{\smallskip}

${\bm C}_{\infty {\rm v}}$&IR&$m$&$R_{\beta}$&$\sigma_{\rm v}$\\ \noalign{\smallskip}\hline\noalign{\smallskip}
&$A_0/B_0$&0 &$1$&$\pm1$\\ \noalign{\smallskip}\hline\noalign{\smallskip}
&$E_m$& $1,2,...$&$\left(
                         \begin{array}{cc}
                           {\rm e}^{{\rm i}\beta m} & 0 \\
                           0 & {\rm e}^{-{\rm i}\beta m} \\
                         \end{array}
                       \right)$
&$\left(\begin{array}{cc}
                           0 & 1 \\
                           1 & 0 \\
                         \end{array}
                       \right)$\\ \noalign{\smallskip}\hline\noalign{\smallskip}
 \end{tabular}
 \end{table}

Due to the negative $g$ factor, the
lowest qubit state $\ket{A_0+}=\ket{A_0}\otimes{\ket{+}}$ is parallel
to the magnetic field direction, while
$\ket{A_0-}=\ket{A_0}\otimes{\ket{-}}$ is the qubit state with spin
projection antiparallel to the magnetic field direction.
The first-order perturbative correction to the qubit states is written as
$\ket{\delta A_0\pm}$. Thus, we can write the SOI corrected qubit states
as $\ket{\uparrow\downarrow}=\ket{A_0\pm}+\ket{\delta A_0\pm}$,
where the normalization factor is omitted as the correction is small.
Correspondingly, the dipole moment is equal to
\begin{eqnarray}\label{dipole}
  {\bf d}=\sum_{j=x,y}\bra{\uparrow}{\bf r}\cdot{\bm e}_{j}\ket{\downarrow}{\bm e}_{j}&=&
\sum_{j=x,y}\Big(\bra{A_0+}{\bf r}\cdot{\bm e}_{j}\ket{\delta A_0-}{\bm e}_{j}\nonumber\\
&&+\bra{\delta A_0+}{\bf r}\cdot{\bm e}_{j}\ket{A_0-}{\bm e}_{j}\Big).
\end{eqnarray}
Since $l_{\rm so}\gg l$, we approximate the unitary operator $U$ with $I_2$,
where $I_2$ is the identity $2\times2$ matrix.
After noticing that $\bra{\pm}s_z\ket{\mp}=-1/2$, $\bra{\pm}s_z\ket{\pm}=0$,
we find the SOI-induced corrections of the qubit states
\begin{eqnarray}\label{popravke}
\ket{\delta A_0\pm}&=&\frac{|g|\mu_BB}{2l_{\rm so}}\sum_{i}\Big(\frac{\bra{E_a^i}x v_x+y v_y\ket{A_0}}{\epsilon^i\pm |g|\mu_B B}\ket{E_a^i\mp}\nonumber\\
&&+\frac{\bra{E_b^i}x v_x+y v_y\ket{A_0}}{\epsilon^i\pm |g|\mu_B B}\ket{E_b^i\mp}\Big).
\end{eqnarray}
Additionally, transition dipole matrix elements are labeled as
\begin{equation}\label{XY}
  X^i=\bra{E_a^i}x\ket{A_0},\;Y^i=\bra{E_a^i}y\ket{A_0}.
\end{equation}
Since the Zeeman splitting is much smaller
than the orbital excitation energies, $|g|\mu_B B\ll \epsilon^i$,
the approximation $\epsilon^i\pm |g|\mu_B B\approx \epsilon^i$ can be made.
Thus, Eq.~\eqref{popravke} is transformed into
\begin{eqnarray}\label{popravke2}
\ket{\delta A_0\pm}=\frac{|g|\mu_BB}{2l_{\rm so}}\sum_{i}&&\Big(\frac{X^i v_x+Y^i v_y}{\epsilon^i}\ket{E_a^i\mp}\nonumber\\
+&&\frac{(X^{i})^* v_x+(Y^i)^* v_y}{\epsilon^i}\ket{E_b^i\mp}\Big),
\end{eqnarray}
where $(X^i)^*$ and $(Y^i)^*$ are the complex conjugates of $X^i$ and $Y^i$, respectively.
Components of the dipole moment can now be written in a more compact form
\begin{eqnarray}
  d_x &=& \frac{2|g|\mu_BB}{l_{\rm so}}\sum_i\frac{|X^i|^2v_x+{\rm Re}(X^i(Y^i)^*)v_y}{\epsilon^i},\nonumber\\
  d_y &=& \frac{2|g|\mu_BB}{l_{\rm so}}\sum_i\frac{|Y^i|^2v_y+{\rm Re}(X^i(Y^i)^*)v_x}{\epsilon^i},\label{dx,dy}
\end{eqnarray}
where ${\rm Re}(X^i(Y^i)^*)$ stands for the real part of $X^i(Y^i)^*$.
Potential dependent parameters that enter Eq.~\eqref{dx,dy} are the transition dipole matrix elements and the excitation energies.
Besides them, dipole moment components are dependent on the spin-orbit angle $\nu$, magnetic field angle $\alpha$, and the
angle $\varphi$ between the [100] crystallographic direction and the $x$ axis.

A further simplification of Eq.~\eqref{dx,dy} stems from the existence of the vertical mirror symmetry $\sigma_{\rm v}$,
requiring that ${\rm Re}(X^i(Y^i)^*)$ must be zero. This can be proven in a few simple steps.
First, we deduce from the matrix of an IR $E_1$, representing the vertical mirror plane, that $\sigma_{\rm v}$ transforms one IR vector into the other, $E_1(\sigma_{\rm v})\ket{E_{a,b}^i}=\ket{E_{b,a}^i}$.
Furthermore, $y$ remains unchanged, while $x$ acquires a minus sign,
leading to the following behavior of the transition matrix elements $X^i$ and $Y^i$
under vertical mirror plane symmetry:
\begin{eqnarray}
X^i\xrightarrow{\sigma_{\rm v}}-(X^i)^*, Y^i\xrightarrow{\sigma_{\rm v}}(Y^i)^*.
\end{eqnarray}
From the previous relations, we conclude that the term ${\rm Re}(X^i(Y^i)^*)$
transforms into $-{\rm Re}(X^i(Y^i)^*)$, meaning that this object does not
obey the symmetry of a system and must vanish.

Additionally, rotational symmetry of a system imposes that
matrix elements $|X^i|^2$ and $|Y^i|^2$ are equal.
This can be concluded from the action of the
rotation $C_n$ for an angle $\beta_n=2\pi/n$ around the $z$ axis,
being the element of the group symmetry.
An element $C_n$ leaves the vector $\ket{A_0}$ unchanged and
adds a phase ${\rm exp}({{\rm i}\beta_n})$ to the vector $\ket{E_a^i}$.
Also, it transforms $x$ and $y$ to $x \cos{\beta_n}+y \sin{\beta_n}$
and $-x\sin{\beta_n}+y\cos{\beta_n}$. Thus, $X^i$ and $Y^i$
are transformed into ${\rm exp}(-{\rm i}\beta_n)(X^i \cos{\beta_n}+Y^i \sin{\beta_n})$
and ${\rm exp}(-{\rm i}\beta_n)(-X^i \sin{\beta_n}+Y^i \cos{\beta_n})$, respectively.
Correspondingly,
\begin{eqnarray}
  &&|X^i|^2\xrightarrow{C_{n}}|X^i|^2\cos^2{\beta_n}+|Y^i|^2\sin^2{\beta_n},\nonumber\\
  &&|Y^i|^2\xrightarrow{C_{n}}|X^i|^2\sin^2{\beta_n}+|Y^i|^2\cos^2{\beta_n},
\end{eqnarray}
where we have neglected the ${\rm Re}(X^i(Y^i)^*)$ term, which was previously proven to equal to zero.
Since $|X^i|^2$ and $|Y^i|^2$ must remain unchanged under the
group symmetry operations, we conclude that the relation
$|X^i|^2=|Y^i|^2$ must hold.
Thus, we have obtained a general relation for the dipole moment in the case
of the potential symmetry ${\bf C}_{n{\rm v}}$ ($n\geq3$):
\begin{equation}\label{REZULT}
  {\bf d}_{\uparrow\downarrow}^{{\bf C}_{n{\rm v}}} = \frac{2|g|\mu_BB}{l_{\rm so}}\Big(\sum_i\frac{|X^i|^2}{\epsilon^i}\Big)(v_x{\bf e}_x+v_y{\bf e}_y).
\end{equation}
In these situations, the absolute value of the dipole moment $|{\bf d}_{\uparrow\downarrow}^{{\bf C}_{n{\rm v}}}|^2\sim(1+\sin{2\alpha}\sin{2\nu})$
is independent of the orientation of the potential with respect to the crystallographic frame.

Analogous analysis can be conducted in the ${\bf C}_{\infty {\rm v}}$ case.
Since the matrix form of the IRs $A_0$ and $E_1$ (see Table~\ref{Cnv}) for this symmetry group is the
same as for ${\bf C}_{n {\rm v}}$, the procedure is exactly the same
if the change $\beta_n\rightarrow\beta$ in the previous discussion is made.

As an example, we implement the derived formula~\eqref{REZULT} in the case of the
isotropic two-dimensional harmonic confinement
$V^{\rm iho}(x,y)=1/2m^*\omega^2(x^2+y^2)$ with ${\bf C}_{\infty {\rm v}}$ symmetry,
assuming only one excited level in the perturbative correction of the qubit states.
With the help of the states $\psi_{0}$ and $\psi_{1}$,
corresponding to the ground and the first excited states of the
one-dimensional harmonic oscillator, we can
define the ground state $\ket{A_0}$ and two complex
conjugate eigenstates $\ket{E_a}$ and $\ket{E_b}$ of the degenerate level:
$\ket{A_0}=\psi_{0}(x)\psi_0(y)$,
$\ket{E_a}=(\psi_{0}(x)\psi_1(y)+{\rm i}\psi_{1}(x)\psi_0(y))/\sqrt{2}$,
and $\ket{E_b}=(\psi_{0}(x)\psi_1(y)-{\rm i}\psi_{1}(x)\psi_0(y))/\sqrt{2}$.
In this case, the squared norm of the transition matrix element is equal to $|X|^2=\hbar/4m^*\omega$.
Using the energy difference of the ground and the first excited energy level $\epsilon=\hbar\omega$
and the confinement length $l=\sqrt{\hbar/m^*\omega}$,
an expression for the dipole moment is obtained~\cite{MSL16}:
\begin{equation}\label{REZULTlho}
  {\bf d}^{\rm iho}_{\uparrow\downarrow} = \frac{|g|\mu_BBm^*l^4}{2l_{\rm so}\hbar^2}\Big(v_x{\bf e}_x+v_y{\bf e}_y\Big).
\end{equation}
\subsection{Dipole moment for systems with ${\bf C}_{2{\rm v}}$ or ${\bf C}_{1{\rm v}}$ symmetry}
As the next step, we discuss potentials with ${\bf C}_{2{\rm v}}$ symmetry.
In this case, coordinates $x$ and $y$ transform according to the IRs $B_1$ and $A_1$, respectively.
Their symmetry behavior imposes the following:
$x(y)$ couples the ground state $\ket{A_0}$
with states transforming according to the
IR $B_1(A_1)$ (see the right-hand panel of FIG.~\ref{SCHEME}). Thus, the SOI-induced corrections of the qubit states are
\begin{eqnarray}\label{popravkeC2v}
\ket{\delta A_0\pm}&=&\frac{|g|\mu_BB}{2l_{\rm so}}\sum_i\Big(\frac{\bra{B_1^i}x v_x\ket{A_0}}{\epsilon_{B_1}^i}\ket{B_1^i\mp}\nonumber\\
&&+\frac{\bra{A_1^i}y v_y\ket{A_0}}{\epsilon_{A_1}^i}\ket{A_1^i\mp}\Big),
\end{eqnarray}
where $\epsilon_{B_1}^i(\epsilon_{A_1}^i)$ is the energy difference between the
energy level transforming according to the IR $B_1(A_1)$ and the ground-state energy.
We define the transition matrix elements as
\begin{equation}\label{XYC2v}
  X^i=\bra{B_1^i}x\ket{A_0},\;Y^i=\bra{A_1^i}y\ket{A_0},
\end{equation}
and obtain the formula for the dipole moment,
\begin{equation}\label{REZULTC2v}
  {\bf d}^{{\bf C}_{2{\rm v}}}_{\uparrow\downarrow} = \frac{|g|\mu_BB}{l_{\rm so}}\sum_i\Big(\frac{|X^i|^2}{\epsilon_{B_1}^i}v_x{\bf e}_x+\frac{|Y^i|^2}{\epsilon_{A_1}^i}v_y{\bf e}_y\Big).
\end{equation}
In this case, anisotropy of the dipole moment appears since it is not forbidden that $\sum_i |X^i|^2/\epsilon_{B_1}^i$
differs from $\sum_i |Y^i|^2/\epsilon_{A_1}^i$.

The anisotropy of the dipole moment can be illuminated using the example of
the anisotropic two-dimensional harmonic potential $V^{\rm aho}(x,y)=1/2m^*(\omega_x^2x^2+\omega_y^2 y^2)$,
with different confinement lengths $l_x=\sqrt{\hbar/m^*\omega_x}$
and $l_y=\sqrt{\hbar/m^*\omega_y}$ along the $x$ and $y$ directions.
We set $l=l_x$ and $l_y=kl$, where $k<1$ is the measure of anisotropy.
We assume two excited orbital states in the perturbative correction: one of type $A_1$ and one of type $B_1$.
In this case we define the ground state $\ket{A_0}=\psi_{0}(x)\psi_0(y)$
and two excited orbital states $\ket{A_1}=\psi_{0}(x)\psi_1(y)$ and $\ket{B_1}=\psi_{1}(x)\psi_0(y)$,
where $\psi_{0/1}(x/y)$ represents the ground or first excited state (subscript 0 or 1, respectively) of
the one-dimensional harmonic oscillator problem in the $x$ or $y$ direction.
The obtained result
\begin{equation}\label{REZULTbho}
  {\bf d}^{\rm aho}_{\uparrow\downarrow} = \frac{|g|\mu_BBm^*l^4}{2l_{\rm so}\hbar^2}\Big(v_x{\bf e}_x+k^4v_y{\bf e}_y\Big)
\end{equation}
is again consistent with Ref.~\cite{MSL16}.

In the case of the ${\bf C}_{1{\rm v}}$ symmetry, using a similar analysis as in the
previous case, we obtain the expression for the dipole moment
\begin{equation}\label{REZULTC2v}
  {\bf d}^{{\bf C}_{1{\rm v}}}_{\uparrow\downarrow} = \frac{|g|\mu_BB}{l_{\rm so}}\sum_i\Big(\frac{|{X^i}|^2}{\epsilon_{B_0}^i}v_x{\bf e}_x+\frac{|Y^i|^2}{\epsilon_{A_0}^i}v_y{\bf e}_y\Big),
\end{equation}
where $X^i=\bra{B_0^i}x\ket{A_0}$, $Y^i=\bra{A_0^i}y\ket{A_0}$,
and $\epsilon_{A_0/B_0}^i$ is the energy difference between the
nondegenerate energy level transforming according to the IR $A_0/B_0$ and the ground-state energy.
\subsection{Dipole moment for systems with ${\bf C}_{n}$ or ${\bf C}_{\infty}$ symmetry}\label{CnSIM}
In the case of ${\bf C}_{n}$ symmetry, all IRs $A_{m}$ ($m\in(-n/2,n/2]$)
are one dimensional and represent an element of symmetry $C_n^s$ $(s=0,1,...,n-1)$
as ${\rm e}^{{\rm i} 2\pi m s/n}$. Besides the geometric symmetry, the time-reversal symmetry $\Theta$ should be included also~\cite{TIMrev}.
Time-reversal $\Theta$ changes the sign of the quantum number $m$ labeling the IR vector $\ket{A_m}$,
since it acts as a complex conjugation in the orbital space:
\begin{equation}
\Theta\ket{A_m}=\ket{A_{-m}}.
\end{equation}
The eigenproblem of the Hamiltonian $H\ket{A_m}=\epsilon_{m}\ket{A_m}$, when
combined with the commutation relation $[\Theta,H_0]=0$, gives us
\begin{equation}
H\ket{A_{-m}}=\epsilon_m\ket{A_{-m}},
\end{equation}
stating that, for $n\geq3$, vectors $\ket{A_{m}}$ and $\ket{A_{-m}}$
are eigenstates of the degenerate level $\epsilon_m$.
To this degenerate level corresponds the reducible representation
$A_m\oplus A_{-m}$ (except for $m=n/2$). The
representation $A_m\oplus A_{-m}$ is equivalent to
the IR $E_m$ of the ${\bf C}_{n{\rm v}}$ group
(see Table~\ref{Cnv}) if the generator $\sigma_{\rm v}$ is neglected.
In other words, Eq.~\eqref{dx,dy} for the dipole moment is valid also in this case,
since it is obtained without assuming the presence of vertical mirror symmetry.
In this case vectors $\ket{E_{a}^i}$ and $\ket{E_{b}^i}$ coincide with
$\ket{A_{1}^i}$ and $\ket{A_{-1}^i}$, respectively.

A further simplification of Eq.~\eqref{dx,dy} appears for systems whose
symmetry element is $\pi/2$ rotation.
This happens if the relation $n=4r$ ($r\in\mathds{N}$) is satisfied.
Since ${\rm Re}(X^i(Y^i)^*)=0$ and $|X^i|^2=|Y^i|^2$ in this case,
Eq.~\eqref{REZULT} is relevant.
Using the same reasoning it can be concluded that Eq.~\eqref{REZULT}
is valid in the ${\bf C}_{\infty}$ case also.

Finally, the dipole moment components for the ${\bf C}_{2}$ symmetry are equal to
\begin{eqnarray}\label{REZULTC2}
  ({\bf d}^{{\bf C}_2}_{\uparrow\downarrow})_x &=& \frac{|g|\mu_BB}{l_{\rm so}}\sum_i \frac{|{X^i}|^2 v_x+{\rm Re}(X^i(Y^i)^*)v_y}{\epsilon_{A_1}^i},\nonumber\\
  ({\bf d}^{{\bf C}_2}_{\uparrow\downarrow})_y &=& \frac{|g|\mu_BB}{l_{\rm so}}\sum_i \frac{|{Y^i}|^2 v_y+{\rm Re}(X^i(Y^i)^*)v_x}{\epsilon_{A_1}^i},
\end{eqnarray}
where $X^i=\bra{A_1^i}x\ket{A_0}$, $Y^i=\bra{A_1^i}y\ket{A_0}$,
and $\epsilon_{A_1}^i$ is the energy difference between the
level transforming according to the IR $A_1$ and the ground-state energy.

To conclude, anisotropy of the potential orientation with respect to the
crystallographic frame
is present in systems without the $\pi/2$ group element ($n\neq 4r$, $r\in\mathds{N}$);
isotropic behavior is present if a rotation for $\pi/2$ is the group element, i.e.,
if $n = 4r$ ($r\in\mathds{N})$ or $n=\infty$.

\section{Applications: infinite-wall equilateral triangle, square, and rectangular potential}\label{examples}
The results presented in the previous Section fully explain the
dependence of the Rabi frequency and spin relaxation rate on the
spin-orbit angle, magnetic field direction, and the relative
orientation of the gating potential with respect to the
crystallographic frame.

\begin{figure}[t]
\includegraphics[width=8.5cm]{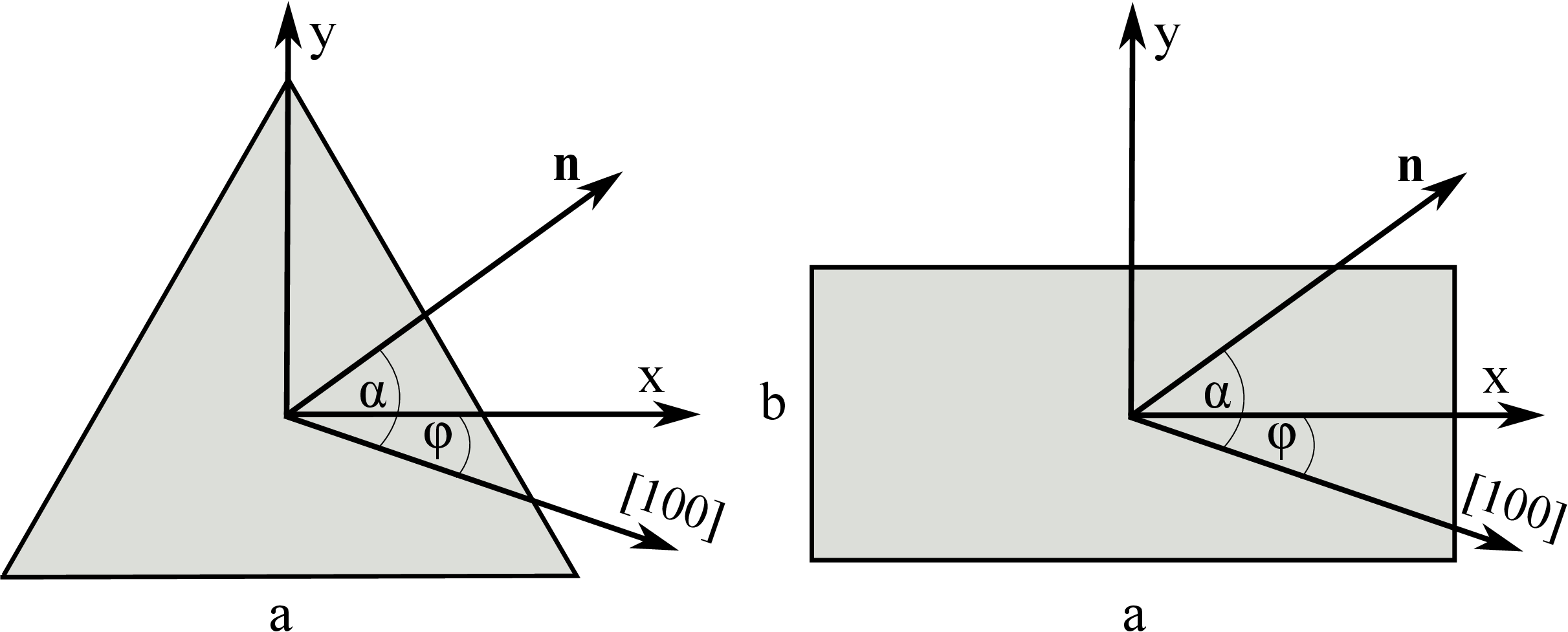}
\caption{\label{TQDRQD}Infinite-wall equilateral triangle (left) and rectangular (right) gating potential.
In both cases, potential is zero inside the area of the polygon;
otherwise it is $\infty$.}
\end{figure}

However, symmetry arguments alone cannot provide us with a
qualitative estimation of the spin relaxation rate,
corresponding to the phonon-allowed spin qubit lifetime.
Since $\Gamma_{\uparrow\downarrow}$ is known for the harmonic gating~\cite{MSL16},
we wish to compare the phonon-induced spin relaxation rate of other
confinement potentials with the known values. To this end, we analyze the spin qubit confined inside the infinite-wall equilateral triangle, square, and rectangular gating potential (see FIG.~\ref{TQDRQD}):
\begin{eqnarray}
  V^{\rm tqd}&=& \left\{
\begin{array}{rl}
0, &  x\in[\frac{y\sqrt{3}-a}3,\frac{a-y\sqrt{3}}3], y\in[\frac{-a\sqrt{3}}6,\frac{a\sqrt{3}}{3}]\;\;\;\label{TQD}\\
\infty,& {\rm otherwise},\\
\end{array} \right.\\
 V^{\rm rqd}&=& \left\{
\begin{array}{rl}
0, &  x\in[-\frac{a}2,\frac{a}2], y\in[-\frac{b}2,\frac{b}2]\\
\infty,& {\rm otherwise}.\\
\end{array} \right.\label{RQD}
\end{eqnarray}

In the first case, Eq.~\eqref{TQD}, the potential has ${\bf C}_{3{\rm v}}$ symmetry and the corresponding
eigenvectors of the spin-independent Hamiltonian $H_0$ transform according to the one-dimensional IRs
$A_0$ and $B_0$ and two-dimensional $E_1$ IR of the ${\bf C}_{3{\rm v}}$ group.
The set of eigenenergies $\epsilon_{p,q}^{\rm tqd}$ and
eigenvectors $\psi^{A_0}_{p,q}$, $\psi^{B_0}_{p,q}$, and $\psi^{E_1\pm}_{p,q}$~\cite{LB85} are
dependent on two parameters $p$ and $q$ that have different sets of allowed values for each IR.
Their concrete form is given in Appendix~\ref{eigen}.

In the second case, Eq.~\eqref{RQD}, the symmetry of the potential is dependent on the ratio $k=b/a\in(0,1]$:
if $k=1$, the symmetry of the problem is ${\bf C}_{4{\rm v}}$; otherwise, ${\bf C}_{2{\rm v}}$ is the symmetry of
the spin-independent Hamiltonian $H_0$. In both situations, eigenenergies and eigenvalues can be found by using
the separation of variables. The set of eigenenergies $\epsilon_{p,q}^{\rm rqd}$ and eigenvectors $\psi_{p,q}^{\rm rqd}$ in this case is
\begin{eqnarray}
\epsilon_{p,q}^{\rm rqd}&=&\frac{\hbar^2\pi^2}{2m^*a^2}(p^2+\frac{q^2}{k^2}),\label{energije}\\
\psi_{p,q}^{\rm rqd}&=&\frac{2}{a\sqrt{k}}\sin{\big[\frac{p\pi}{a}(x+\frac{a}2)\big]}\sin{\big[\frac{q\pi}{ak}(y+\frac{k a}2)\big]},\;\;\;
\end{eqnarray}
defined using the two independent parameters $p\geq1$ and $q\geq1$ that take integer values.
However, these solutions do not have any definite symmetry~\cite{CCM+05}. Therefore, they need to be
symmetrized to apply the general results from Section~\ref{Dipole}.
Symmetry-adapted eigenfunctions can be found in Appendix~\ref{eigenRQD}.

After calculating the transition dipole matrix element
and the excitation energies for two excited states in the perturbative correction~\cite{check}, we obtain the desired results
\begin{eqnarray}\label{v}
  {\bf d}_{\uparrow\downarrow}^{\rm tqd}&=&\frac{3^{24}}{2^{26}35^2\pi^8}\frac{|g|\mu_B B m^*a^4}{l_{\rm so}\hbar^2}(v_x{\bm e}_x+v_y{\bm e}_y),\label{edsrTQD}\\
  {\bf d}_{\uparrow\downarrow}^{\rm rqd}&=&\frac{2^{9}}{3^5\pi^6}\frac{|g|\mu_B B m^*a^4}{l_{\rm so}\hbar^2}(v_x{\bm e}_x+k^4v_y{\bm e}_y),\label{edsrRQD}
  \end{eqnarray}
where the first result corresponds to the infinite-wall equilateral triangle potential,
while the second one is valid for both the infinite-wall square, $k=1$, and rectangular, $k\neq1$, potentials.
Dipole moment constants  $3^{24}/2^{26}35^2\pi^8\approx 3.6\times10^{-4}$
and $2^{9}/3^5\pi^6\approx 2.2\times10^{-3}$ from Eqs.~(\ref{edsrTQD}) and (\ref{edsrRQD})
suggest a much weaker dipole moment when compared to the harmonic gating of the same confinement length
[see Eqs.~\eqref{REZULTlho} and \eqref{REZULTbho}].

Using the relation $\Gamma_{\uparrow\downarrow}\approx|{\bf d}_{\uparrow\downarrow}|^2$,
we conclude that square and rectangular confined QDs
have a relaxation rate that is four orders of magnitude weaker than
the harmonic potential;
in the equilateral triangle case, a decrease of almost six orders of magnitude is observed.
Thus, our result indicates a significant influence of the gating potential
on the spin qubit lifetime and a beneficial role of the equilateral triangle confinement.
\section{Conclusions}\label{Section4}
We have investigated the influence of the gating potential
symmetry on the Rabi frequency and phonon-induced spin
relaxation rate in a single-electron GaAs quantum dot.
Our results suggest that, independently of the symmetry of the gating potential,
both the Rabi frequency and spin relaxation rate are dependent on the
orientation of the magnetic field and the spin-orbit angle.
Additionally, in systems with ${\bf C}_{1{\rm v}}$, ${\bf C}_{2{\rm v}}$, and
${\bf C}_{n}$ ($n\neq 4r$) symmetry, orientation of the quantum dot potential
with respect to the crystallographic reference frame is another
degree of freedom that can be used to tune the desired properties of
the system. The validity of the approach is confirmed on the
known results for the isotropic and anisotropic harmonic potential.
Additionally, we have compared the spin qubit lifetime in
the case of an infinite-wall rectangular, square and equilateral triangle
gating with the harmonic confinement.
Our results  indicate the enhanced lifetime of the spin qubit,
reaching an almost six-orders-of-magnitude increase
in the case of the equilateral triangle gating.
In the end, we emphasize that in the regime of strong electric field, nonlinear effects~\cite{KGS12,RGP15,VP18} cannot be fully explained by the symmetry of the gating potential, thus placing the conclusions of our work in the weak driving regime solely.

\acknowledgments
We thank Nenad Vukmirovi\'{c} for fruitful discussion.
This research is funded by the Serbian Ministry of Science (Project ON171035)
and the National Scholarship Programme of the Slovak Republic (ID 28226).

\appendix
\section{Particle in the infinite-wall equilateral triangle potential: eigenenergies and eigenvectors}\label{eigen}
Here we summarize the results from Ref.~\cite{LB85}
regarding the Schr\"{o}dinger equation solution of the particle
in the infinite-wall equilateral triangle potential, having ${\bf C}_{3{\rm v}}$ symmetry.
Due to the symmetry, eigenvectors
transform according to the one-dimensional IRs $A_0$ and $B_0$ and the two-dimensional IR $E_1$.
The concrete forms of eigenenergies and eigenstates,
\begin{equation}
\epsilon_{p,q}^{\rm tqd}=\frac{8\hbar^2\pi^2}{3m^*a^2}(p^2+pq+q^2),\label{En}
\end{equation}
\begin{eqnarray}
\psi^{A_0}_{p,q}(x,y)&=&\cos{\big[\frac{2\pi q}{a}x\big]}\sin{\big[\frac{2\pi (2p+q)}{a\sqrt{3}}y\big]}\nonumber\\
&&-\cos{\big[\frac{2\pi p}{a}x\big]}\sin{\big[\frac{2\pi (p+2q)}{a\sqrt{3}}y\big]}\nonumber\\
&&-\cos{\big[\frac{2\pi (p+q)}{a}x\big]}\sin{\big[\frac{2\pi (p-q)}{a\sqrt{3}}y\big]},\nonumber\\
q&=&0,1,2,...,\;\;\;p=q+1,q+2,...\label{A1stanje}
\end{eqnarray}
\begin{eqnarray}
\psi^{B_0}_{p,q}(x,y)&=&\sin{\big[\frac{2\pi q}{a}x\big]}\sin{\big[\frac{2\pi (2p+q)}{a\sqrt{3}}y\big]}\nonumber\\
&&-\sin{\big[\frac{2\pi p}{a}x\big]}\sin{\big[\frac{2\pi (p+2q)}{a\sqrt{3}}y\big]}\nonumber\\
&&+\sin{\big[\frac{2\pi (p+q)}{a}x\big]}\sin{\big[\frac{2\pi (p-q)}{a\sqrt{3}}y\big]},\nonumber\\
q&=&1,2,3,...,\;\;\;p=q+1,q+2,...\label{A2stanje}
\end{eqnarray}
\begin{eqnarray}
\psi^{E_1\pm}_{p,q}(x,y)&=&\psi^{B_0}_{p,q}(x,y)\pm{\rm i}\psi^{A_0}_{p,q}(x,y),\nonumber\\
q&=&\frac{1}3,\frac{2}3,\frac{4}3,\frac{5}3,...,\;\;\;p=q+1,q+2,...\;\;\label{E1stanje}
\end{eqnarray}
are dependent on two parameters $p$ and $q$ that have different allowed values for each IR.
\begin{figure}[t]
\centering
\includegraphics[width=4.2cm]{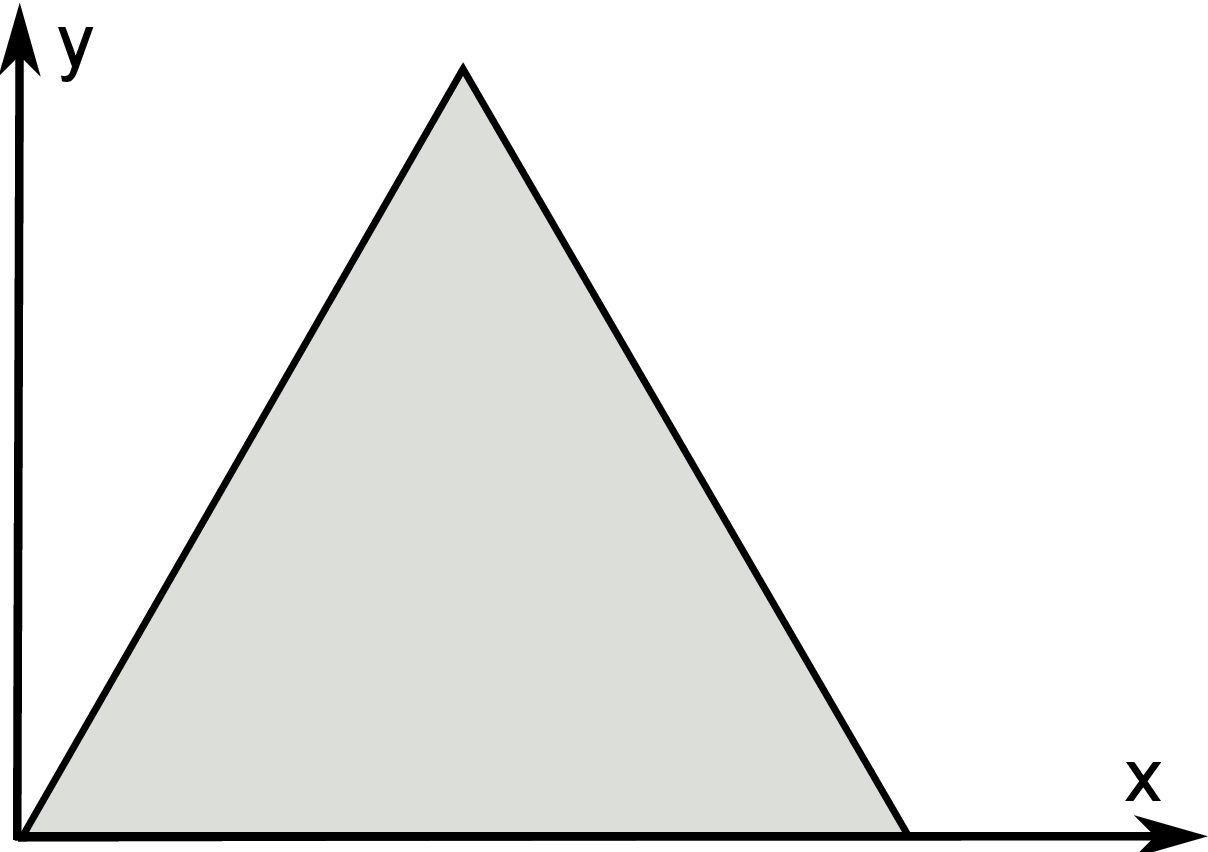}
\caption{\label{TriPot}
Infinite-wall equilateral triangle potential with the point group symmetry ${\bf C}_{3{\rm v}}$. Inside the equilateral triangle potential is 0, otherwise it is $\infty$.}
\end{figure}
Note that the coordinate frame used to derive the previous equations (see FIG.~\ref{TriPot})
differs from the frame used in our work (see the left-hand panel of FIG.~\ref{TQDRQD}). To adapt the eigenfunction from
Eqs.~(\ref{A1stanje})-(\ref{E1stanje}) to our case, a suitable change of coordinates $x\rightarrow x+a/2$
and $y\rightarrow y+a\sqrt{3}/6$ should be made.

\phantom{xxxxxxxxx}

\section{Particle in the infinite-wall square and rectangular potential: eigenvectors}\label{eigenRQD}
The infinite-wall square potential has ${\bf C}_{4{\rm v}}$ symmetry with the corresponding IRs $A_0/B_0$, $A_2/B_2$, and $E_1$.
Eigenvectors that transform according to the given IRs and the set of allowed quantum numbers are
\begin{eqnarray}
\psi^{A_0}_{p,q}(x,y)&=&\cos{\big[\frac{p\pi}{a}x\big]}\cos{\big[\frac{q\pi}{a}y\big]}+\cos{\big[\frac{q\pi}{a}x\big]}\cos{\big[\frac{p\pi}{a}y\big]}\nonumber\\
q&=&1,3,5,...,\;\;\;p=q,q+2,q+4,...\label{A0stanje}
\end{eqnarray}
\begin{eqnarray}
\psi^{B_0}_{p,q}(x,y)&=&\sin{\big[\frac{p\pi}{a}x\big]}\sin{\big[\frac{q\pi}{a}y\big]}-\sin{\big[\frac{q\pi}{a}x\big]}\sin{\big[\frac{p\pi}{a}y\big]}\nonumber\\
q&=&2,4,6,...,\;\;\;p=q+2,q+4,...\label{B0stanje}
\end{eqnarray}
\begin{eqnarray}
\psi^{A_2}_{p,q}(x,y)&=&\cos{\big[\frac{p\pi}{a}x\big]}\cos{\big[\frac{q\pi}{a}y\big]}-\cos{\big[\frac{q\pi}{a}x\big]}\cos{\big[\frac{p\pi}{a}y\big]}\nonumber\\
q&=&1,3,5,...,\;\;\;p=q+2,q+4,...\label{A2stanje}
\end{eqnarray}
\begin{eqnarray}
\psi^{B_2}_{p,q}(x,y)&=&\sin{\big[\frac{p\pi}{a}x\big]}\sin{\big[\frac{q\pi}{a}y\big]}+\sin{\big[\frac{q\pi}{a}x\big]}\sin{\big[\frac{p\pi}{a}y\big]}\nonumber\\
q&=&2,4,6,...,\;\;\;p=q,q+2,q+4,...\label{B2stanje}
\end{eqnarray}
\phantom{xxxxxxxxxxxxxxxxxxxxxxxxxx}
\begin{eqnarray}
\psi^{E_1\pm}_{p,q}(x,y)&=&\cos{\big[\frac{p\pi}{a}x\big]}\sin{\big[\frac{q\pi}{a}y\big]}\pm{\rm i}\sin{\big[\frac{q\pi}{a}x\big]}\cos{\big[\frac{p\pi}{a}y\big]}\nonumber\\
p&=&1,3,5,...,\;\;\;q=p+1,p+3,...\label{Epmstanje}
\end{eqnarray}
In the case of the infinite-wall rectangular potential ${\bf C}_{2{\rm v}}$ symmetry is relevant.
Eigenfunctions transforming according to the IRs $A_0/B_0$ and $A_1/B_1$ and the corresponding set of
quantum numbers are
\begin{eqnarray}
\psi^{A_0}_{p,q}(x,y)&=&\cos{\big[\frac{p\pi}{a}x\big]}\cos{\big[\frac{q\pi}{ka}y\big]}\nonumber\\
q&=&1,3,5,...,\;\;\;p=q,q+2,q+4,...\label{A0C2v}
\end{eqnarray}
\begin{eqnarray}
\psi^{B_0}_{p,q}(x,y)&=&\sin{\big[\frac{p\pi}{a}x\big]}\sin{\big[\frac{q\pi}{ka}y\big]}\nonumber\\
q&=&2,4,6,...,\;\;\;p=q,q+2,q+4,...\label{B0C2v}
\end{eqnarray}
\begin{eqnarray}
\psi^{A_1}_{p,q}(x,y)&=&\cos{\big[\frac{p\pi}{a}x\big]}\sin{\big[\frac{q\pi}{ka}y\big]}\nonumber\\
p&=&1,3,5,...,\;\;\;q=p+1,p+3,p+5,...\;\;\;\label{A0C2v}
\end{eqnarray}
\begin{eqnarray}
\psi^{B_1}_{p,q}(x,y)&=&\sin{\big[\frac{p\pi}{a}x\big]}\cos{\big[\frac{q\pi}{ka}y\big]}\nonumber\\
q&=&1,3,5,...,\;\;\;p=q+1,q+3,q+5,...\;\;\;\label{A0C2v}
\end{eqnarray}
In both cases, eigenenergies are given in Eq.~\eqref{energije} ($k=1$ in the ${\bf C}_{4{\rm v}}$ case and $k\neq1$ for the ${\bf C}_{2{\rm v}}$ symmetry).


\begin{thebibliography}{50}
\bibitem{check} We have explicitly checked that other states appearing in the perturbative expansion can be safely ignored.
\end{thebibliography}
\end{document}